\documentclass[aps,prd,showpacs,groupedaddress,floatfix,nofootinbib]{revtex4}
\usepackage{amsmath}
\usepackage{graphicx}
\usepackage{color}
\usepackage{graphicx}
\usepackage{epsfig}
\usepackage{amsfonts}
\usepackage{amstext}
\usepackage{amssymb}
\newcommand{\beq}{\begin{eqnarray}}
\newcommand{\eeq}{\end{eqnarray}}

\begin{document}

\title{{Reply to ``Comment on `Quantization of FRW spacetimes \\ in the presence of a cosmological constant and radiation' '' }}
\author{N. A. Lemos}
\email{nivaldo@if.uff.br}
\affiliation{Instituto de F\'{\i}sica, Universidade Federal Fluminense,\\
R. Gal. Milton Tavares de Souza s/n$^o$, CEP 24210-340, Niter\'oi, RJ, Brazil}
\author{G. A. Monerat}
\email{monerat@fat.uerj.br}
\affiliation{Departamento de Matem\'atica e Computa\c c\~ao, Faculdade de Tecnologia, Universidade do Estado do Rio de Janeiro, \\
Estrada Resende-Riachuelo s/n$^o$, CEP 27523-000, Resende, RJ, Brazil}
\author{E. V. Corr\^ea Silva}
\email{eduardo.vasquez@pesquisador.cnpq.br}
\affiliation{Departamento de Matem\'atica e Computa\c c\~ao, Faculdade de Tecnologia, Universidade do Estado do Rio de Janeiro, \\
Estrada Resende-Riachuelo s/n$^o$, CEP 27523-000, Resende, RJ, Brazil}
\author{G. Oliveira-Neto}
\email{gilneto@fat.uerj.br}
\affiliation{Departamento de Matem\'atica e Computa\c c\~ao, Faculdade de Tecnologia, Universidade do Estado do Rio de Janeiro, \\
Estrada Resende-Riachuelo s/n$^o$, CEP 27523-000, Resende, RJ, Brazil}
\author{L. G. Ferreira Filho}
\email{gonzaga@fat.uerj.br}
\affiliation{Departamento de Mec\^anica e Energia, Faculdade de Tecnologia, Universidade do Estado do Rio de Janeiro, \\
Estrada Resende-Riachuelo s/n$^o$, CEP 27523-000, Resende, RJ, Brazil}

\begin{abstract}
 The  Comment by Amore {\it et al.} [gr-qc/0611029] contains a valid criticism of the numerical precision of the results reported in a recent   paper of ours [Phys. Rev. D {\bf 73}, 044022 (2006)], as well as fresh ideas on how to characterize a quantum cosmological singularity. However, we argue that, contrary to what is suggested in the  Comment, the quantum cosmological models
we studied show hardly any sign of singular behavior.


\end{abstract}



\maketitle

The previous Comment \cite{Amore} presents a criticism of  the numerical precision  and the physical interpretation of our recent results on the quantum dynamics of
Friedmann-Robertson-Walker cosmological models with radiation and a negative cosmological constant \cite{Monerat}.

 Concerning the accuracy of the numerical results the criticism is justified, although this has played no decisive  role in our  analysis, as acknowledged by the authors of the Comment \cite{Amore}. The best numerical results for the energy levels that we obtained  by means of  the Chhajlany-Malnev method  are indeed correct up to 20 digits, but the worst are correct only up to 5 digits.
 The lowest precision typically occurs for the intermediate energies in the spectrum. The need to deal with a modified potential instead of the exact one, inherent in the Chhajlany-Malnev method, gives rise to significant errors, particularly in the case $k=-1$.
 Therefore,
  the numerical method that we used  is not uniformly precise, contrary to what is suggested in our paper. The ``Variational Sinc Collocation Method'' employed by  Amore {\it et al.} is more uniformly precise and appears to converge more rapidly than the method chosen by us. Thus, the powerful numerical tool used in Ref. \cite{Amore}
will probably  be useful for future research in quantum cosmology.

  As to the physical interpretation, we have some points of disagreement with the authors of the Comment. They criticize us for basing our conclusions in regard to  singularity avoidance solely on the behavior of the expectation value of the scale factor, and advance some interesting ideas on singularity identification that deserve a closer scrutiny. Firstly, they propose a comparison between the classical and quantum probabilities for finding the scale factor near $a=0$.
In the cases $k=0$ or $k=1$ the classical probability of finding the scale factor in the interval $(0,\epsilon )$ is given by
\begin{equation}
P_{cl}(a<\epsilon )=N_{cl}\int_0^{\epsilon} \frac{da}{\sqrt{{\cal E} - V_e(a)}} \approx  \frac{N_{cl}} {\sqrt{{\cal E}}}\, \epsilon
\end{equation}
for $\epsilon $ small enough, since the potential $V_e(a)$ vanishes at $a=0$. As to the quantum probability, one has
\begin{equation}
P_{q}(a<\epsilon )=\int_0^{\epsilon} \vert \Psi (a,t)  \vert^2 da \approx   \vert \Psi (\epsilon ,t) \vert^2\, \epsilon\, .
\end{equation}
Since $N_{cl}$ and $\cal E$ are both independent of $\epsilon$,
it follows that
\begin{equation}
\frac{P_{q}(a<\epsilon )}{P_{cl}(a<\epsilon )} \propto  \vert \Psi (\epsilon ,t) \vert^2 \to 0 \,\,\, {\rm as} \,\,\, \epsilon \to 0
\end{equation}
because the wave functions employed by us vanish at $a=0$. Therefore, it is impossible for the quantum probability to be greater than the classical probability
in a sufficiently small neighborhood of the classical singularity. The authors of the Comment claim that this is merely ``a consequence of the Dirichlet boundary conditions on the wave function''. However, the boundary condition $\Psi (a=0, t)=0$
is a special case of the general boundary condition \cite{Reed}
\begin{equation}
\Psi^{\prime}(0,t) = \alpha \Psi (0,t)\, , \,\,\,\,\, \alpha \in \mathbb{R}\cup \{ \infty \}\, ,
\end{equation}
which is necessary to make sure that  the Hamiltonian operator be self-adjoint (the prime denotes derivative with respect to $a$). This includes wave functions such that both $\Psi (a,t)$ and $\Psi^{\prime}(a,t)$ vanish at $a=0$. In these cases,  as $\epsilon \to 0$ the quantum probability $P_{q}(a<\epsilon )$ tends to zero even faster as compared to the   classical probability $P_{cl}(a<\epsilon )$. Thus, there is a large  set of states
for which the quantum probability of finding the scale factor near $a=0$ is arbitrarily smaller that the corresponding classical probability. Taking $P_q(a<\epsilon ) > P_{cl}(a<\epsilon )$ for all sufficiently small $\epsilon$ as a singularity indicator, an idea further discussed below, it is clear that there is a large set of nonsingular quantum states. Of course, this does not prove that {\it all} states are nonsingular, which prevents us from drawing any sharp conclusion.

The case $k=-1$ requires some qualifications. The effective potential in this case is
\begin{equation}
V_e(a) = 12 \vert \Lambda \vert a^4 - 36 a^2\, ,
\end{equation}
and a simple sketch of the graph of $V_e$ shows that for positive energies there is no classically forbidden region around $a=0$. Thus, the statement of Amore {\it et al.} \cite{Amore} that for $k=-1$ the classical probability vanishes in a sufficiently small neighborhood of $a=0$ is correct only if the energy ${\cal E}=12E$ is negative. One can easily construct wave packets with positive mean energy which invalidate the arguments presented in the Comment in the case $k=-1$ as well.

The singularity criterion $P_q(a<\epsilon ) > P_{cl}(a<\epsilon )$ suggested by our critics
is a welcome attempt to improve our means of recognizing a quantum cosmological singularity.
Unfortunately, it is not exempt from difficulties. If $k=-1$ and the energy is negative the classical model is nonsingular ($a=0$ cannot be reached) and $P_{cl}(a<\epsilon )=0$ for $\epsilon$ sufficiently small. The fact that $P_q(a<\epsilon ) > P_{cl}(a<\epsilon )$ just because $P_{cl}(a<\epsilon )=0$ should not be taken by itself as a symptom that the quantum model is singular, as our critics seem to do.
Since classically forbidden regions are quantum mechanically allowed, this criterion would force us to conclude that all  nonsingular classical cosmological models are singular at the quantum level. This would mean that, instead of contributing to the avoidance of singularities, the quantization process creates singularities where classically there was none, which does not sound very sensible to us.

According to the authors of the preceding Comment,  the study of the behavior of the expectation value of the scale factor alone is not sufficient to draw sharp conclusions on the presence or absence of singularities. Although they may be right, we cannot fail to remark that the expectation value of the scale factor has often been used  as a singularity indicator  in quantum cosmology \cite{Lund}.
Our critics assert that it is more appropriate to study  the behavior of the probability of finding the scale factor below a given value. Intuitively, if $P_q(a<\epsilon )$ ever became  close to one for very small $\epsilon$ the probability density $\vert \Psi (a,t)\vert^2$ would be sharply peaked near $a=0$, and   we would have  an indication of a singularity. But then most likely $\langle a \rangle$ would be very small too, so that $P_q(a<\epsilon )$ appears to be  a singularity indicator essentially equivalent to $\langle a \rangle$.
Both our analysis and the more accurate numerical study of Amore {et al.} \cite{Amore}  of the dynamical evolution of the models give no hint that such an extreme condition is reached of very small   $\langle a \rangle$
or $P_{q}(a<\epsilon )$ close to unity for very small $\epsilon$ .
This behavior of the wave packets is reflected --- not so faithfully as we would like, we concede --- in the expectation value of the scale factor, and provides evidence that the quantum models do not develop singularities.
Our critics  state correctly  that the expectation value of the scale factor can never vanish, but this does not mean that ``hardly any conclusion can be drawn from the study of $\langle a \rangle$". Our conjecture  that $P_q(a<\epsilon )$ and   $\langle a \rangle$ are essentially equivalent  singularity indicators should not be mistaken for a rigorous statement, and  the proposal of the study of  $P_q(a<\epsilon )$ as a more reliable singularity indicator deserves further investigation.

As a matter of fact,
it is hard
 to make clear-cut statements on presence or absence of cosmological  singularities  in the quantum regime because there is no general
agreement on  what constitutes a quantum singularity.  This is one of the fundamental problems of quantum cosmology: there is no indisputable definition of a quantum singularity. Owing to the lack of a general, physically reasonable and mathematically precise characterization of a quantum singularity, all statements on existence or nonexistence of singularities in quantum cosmological models are open to criticism, and this is clearly recognized in the conclusions of our paper \cite{Monerat}. The suggestion of the authors of the  Comment that one should look at  the  probability $P_{q}(a<\epsilon )$ is interesting and  worthy of a deeper analysis, although we suspect  that it   also fails to provide an indubitable singularity indicator.

   In short, our critics are right in considering that the evidence we provided was  not sufficiently convincing to show beyond  doubt that the quantum cosmological models we studied are nonsingular.
We hope we have supplied  more satisfactory arguments in this Reply. Regrettably, to our knowledge
sharp and conclusive arguments were not available when we wrote our paper, and remain so. At any rate, the suggestions made in the Comment
seem a valuable effort towards a better understanding of quantum cosmological singularities.

\end{document}